# A Study of the Three-Body Force Effect on the EOS Properties of Asymmetric Nuclear Matter

Mansour H. M. M.*[1], M. El-Zohry[2] and A. E. Elmeshneb[2]

[1] *Physics Department, Faculty of Science, Cairo University, Cairo 12613, Egypt*

[2] *Physics Department, Faculty of Science, Sohag University, Sohag 82524, Egypt*

*E-mail: mansourhesham@yahoo.com

**Abstract.** Three-body effects are studied for both asymmetric nuclear matter and pure neutron matter to calculate the nuclear Equation of State (EOS). The Brueckner-Hartree-Fock (BHF) approximation is used using CD BonnB and Argonne V18 potentials. A two-body density dependent Skyrme potential is added to reproduce the empirical saturation point. Good agreement is obtained in comparison with exact calculation including three-body forces.

## 1. Introduction

The nuclear matter has attracted the attention for many years because of its relevance to astrophysics and heavy ion collisions. For symmetric nuclear matter, it is a hypothetical system of equal number of protons and neutrons ($\rho_n = \rho_p$), where $\rho_n$ represents the neutron density and $\rho_p$ represents the proton density, but in case of asymmetric nuclear matter ($\rho_n \neq \rho_p$). To distinguish between symmetric and asymmetric nuclear matter we have to use a symmetry parameter $\alpha$ which equal $(\rho_n - \rho_p)/\rho$, in case of symmetric nuclear matter $\alpha = 0$ and $\alpha = 1$ for pure neutron matter, where $\rho$ is the total density equals to $(\rho_n + \rho_p) = \frac{2k_f^3}{3\pi^2}$. Here $k_f$ represents the Fermi momentum. This physical quantity E/A as a function of the nucleon density is called the Equation of State (EOS). The EOS is a thermo dynamical equation describing a mathematical relationship between state variables of matter in equilibrium under a given set of physical conditions. Plotting the *E/A* with the density gives us a curve, which decreases with increasing density to reach the minimum value point called saturation point. The exact experimental value of saturation point $\rho_0 = 0.16 \pm 0.01 fm^{-3}$ at $E/A = -16 \pm 1 MeV$ [1].

By using the properties of the nuclear medium in an extremely wide range of density, the internal structure of neutron stars can be determined. Some neutron stars rotate very rabidly up to 716 times per second or approximately 43000 revolutions per minute [2, 3] giving a linear speed at the surface on the order 165*c* and emit beams of electromagnetic radiation as pulsars. $10^{11} \to 10^{12} K$. Indeed, the discovery of pulsars in 1967 first suggested that neutron stars exist.

The temperature inside neutron star is $10^{11} \rightarrow 10^{12} K$. The pressure increases from $3 \times 10^{33} \rightarrow 1.6 \times 10^{35} Pa$ from the inner crust to the center. The EOS of neutron star still not known, it is assumed that it differs from that of white dwarf. Unfortunately, the relation between density and mass is not fully known, which means that we have uncertainties in radius.

We used in this work the CD-Bonn potential, but unfortunately, this potential cannot be used directly in perturbative calculations and should be normalized firstly. This potential is based on meson exchange, all mesons with masses below the nucleon mass like $(\pi, \eta, \rho(770); \omega(782))$ were included.

One of the most accurate *NN* interactions is the Argonne *V*18 potential [6]. It is constructed by a set of two-body operators, which arise naturally in meson exchange processes, but the form factors are partly phenomenological (except, of course the one-pion exchange). Actually the *NN* potential are constructed by fitting np data for $T = 0$ states and either np or pp data for $T = 1$ states. One of the examples for potentials fit to np in all states are Argon V14 [7], Urbana *V14* [8] and most of the Bonn potentials [9], unfortunately, potential models which fit only the np data often give a poor description of pp data [10], even after applying the necessary Coulomb correction. On the other hand potentials fit to pp data in $T = 1$ states five only medium description of np data. Fundamentally, this problem is due to charge-independence breaking in the strong interaction. Argonne *V*18 potential is a high quality, non-relativistic, local *NN* interaction with explicit charge dependence and charge dependence and charge asymmetry.

## 2. Theoretical approaches

Different approaches have been taken for the investigation of the properties of Nuclear matter. Brueckner reaction matrix $G$ represents the starting point in the BBG theory [4, 5]. The BBG theory is based on a linked cluster expansion of the energy per nucleon of nuclear matter, which in the case of asymmetric nuclear matter depends on the isospin components of the two colliding nucleons. Then the $G$ matrix can be represented as:

$$G(\rho, \beta, \omega) = U_{NN} + \frac{U_{NN}}{V} \sum_{k_1 k_2} \frac{|K_1 K_2 > Q(K_1 K_2) < K_1 K_2|}{\omega - \epsilon(K_1) - \epsilon(K_2) + i\eta} G(\rho, \beta, \omega), \qquad (1)$$

where $V$ is the volume of the system. The single-particle momentum denoted by $K_i$. $U_{NN}$ is the *NN* interaction, and $\omega$ is the starting energy where $\beta$ represents the asymmetry parameter is defined as $\beta = (\rho_n - \rho_p)/\rho$. The operator $Q(K_1, K_2)$ projects on intermediate scattering states in which both nucleons are above the Fermi sea (Pauli operator). Our calculations are performed with a variety of nucleon-nucleon potentials, among which is the $AV18$ potential [6] and the BonnB potential [11] for the two-body nuclear force, and their corresponding potentials for the three-body force. The single particle energy is defined as:

$$\epsilon(k) = \frac{\hbar^2}{2m}k^2 + U_{BHF}(k) \qquad (2)$$

The Brueckner–Hartree–Fock (BHF) approximation for the s.p. potential $U(k,n)$, using the continuous choice [12] is

$$U_{BHF}(k) = \frac{1}{V}\sum_{k'} n(k')\,\mathrm{Re}\langle kk'|G(\epsilon(k)+\epsilon(k'))|kk'\rangle_A \qquad (3)$$

where the subscript "$A$" indicates antisymmetrization of the matrix element. In this approach equations (1)–(3) have to be solved self consistently. In the BHF approximation the energy per nucleon as a function of density $\rho$ and the isospin asymmetry $\beta$ is:

$$E_A(\rho,\beta) = \frac{3}{5}\frac{\hbar^2}{2m}\left[\left(\frac{1-\beta}{2}\right)^{5/3} + \left(\frac{1+\beta}{2}\right)^{5/3}\right](3\pi^2\rho)^{2/3} + \frac{1}{2V}\mathrm{Re}\sum_{k,k'} n(k)n(k')\langle kk'|G[\rho,\beta;\epsilon(k)+\epsilon(k')]|kk'\rangle \qquad (4)$$

The only quantity needed to solve the Beta-Goldstone equation is the bare $NN$ interaction $U_{NN}$. In fact, we would like to mention about the historical story about the choice of the forces between the nucleons, since all the results for the nuclear Equation of State (EOS) using the realistic two-body force fail to reproduce the correct saturation point extracted from the experimental data on a wide set of nuclei. Despite of many different two-body forces have been used in order to reproduce the different saturation points, but, unfortunately, all of the points are outside the experimental constraints. In the case of Brueckner–Hartree–Fock (BHF) approximation with standard choice of the single particle potential, all the saturation points appear to lie along the so called "coester band" [13]. By using the continuous choice, the "coester band" reproduces closely results in a more limited region of the energy-density plane. However, the empirical region is missed [14]. Moreover, the use of the three-hole line contribution according to the Bethe-Brueckner-Goldstone (BBG) expansion does not reproduce correctly the saturation points since the contribution appear to be quite small [15, 16].

The variation method has been used [17] to calculate the various properties of hot and frozen homogeneous fermionic fluids such as symmetric and asymmetric nuclear matter [18]. One of the models of the variational methods in the Lowest Order Constrained Variational (LOCV) approximation has been studied and compared with the Brueckner-Hartree-Fock (BHF) approach and it is shown under the same conditions and approximations that the two approaches are equivalent [19]. Finally, the results show the same conclusion without achieving the empirical saturation point.

### 3. Three body force

The fail of reproducing the empirical saturation points requires the needs of including three body force (TBF). However, it seems to be very difficult to reproduce the experimental binding energies of light nuclei and correct saturation points using one simple set of the TBF [20]. However, many progresses have been made during last decades in the theory of TBF. Recently, different models have been used. Two major lines of this approach have been pursued in the past, the first one involving a semi phenomenological determination of the TBF [22, 23] such as Urbana model. The Urbana model consists of an attractive term $V_{ijk}^{2\pi}$ due to two-pions exchange with excitation of an intermediate Δ-resonance, and a repulsive phenomenological central term $V_{ijk}^{R}$

$$V_{ijk} = V_{ijk}^{2\pi} + V_{ijk}^{R}, \tag{5}$$

which is the attractive part contribution with a cyclic sum over the nucleon indices $i, j, k$ of products of anticommutator {,} and commutator [,] terms

$$V_{ijk}^{2\pi} = A \sum_{cyc} \left( \{X_{ij}, X_{jk}\}\{\tau_i, \tau_j, \tau_j \cdot \tau_k\} + \frac{1}{4}[X_{ij}, X_{jk}][\tau_i, \tau_j, \tau_j \cdot \tau_k] \right), \tag{6}$$

where

$$X_{ij} = Y(r_{ij})\sigma_i \cdot \sigma_j + T(r_{ij})S_{ij} \tag{7}$$

is the one-pion exchange operator, $\sigma$ and $\tau$ are the Pauli spin and isospin operators, and $S_{ij} = 3\left[(\sigma_i r_{ij})(\sigma_j r_{ij}) - \sigma_i \sigma_j\right]$ is the tensor operator. $Y(r)$ and $T(r)$ are the Yukawa and tensor functions, respectively, associated to the one–pion exchange, as in the two–body potential.

The repulsive part represented as:

$$V_{ijk}^{R} = U \sum_{cyc} T^2(r_{ij}) T^2(r_{jk}), \tag{8}$$

where the constants A and U in the BHF can be adjusted to reproduce the observed nuclear properties. For example in Ref. 24 found $A = -0.0333$ and $U = 0.0038$ by fitting properties of light nuclei ($^3H$, $^4He$), while in the Urbana TBF using the variational approach, the two parameters A and U can be adjusted to reproduce the observed ($^3H$, $^4He$) binding energy only, but unfortunately these two parameters fail to reproduce the correct saturation point of symmetric nuclear matter [20, 21]. It is worth to be mentioned that the TBF parameters values are quite different in both BHF calculations and variational approach, and it is found that the repulsive term in the BHF approach is much weaker [25].

It contains the contribution due to the medium modification of the two-meson ($\pi\pi, \pi\rho, \rho\rho$) exchange part of the nucleon-nucleon (*NN*) interaction, the $\rho\pi\gamma$ diagram and the contribution associated to the $\varphi$ and $\omega$ meson exchange. The effect of the TBF has been included in the calculations along the same line as in Ref. 32, where it is reduced to an effective two-body interaction in order to avoid the difficulty of the full three-body problem. We can write down the equivalent two-body potential, which is given in r-space by

$$\langle \vec{r}_1 \vec{r}_2 | V_3 | \vec{r}_1' \vec{r}_2' \rangle = \frac{1}{4} Tr \sum_n \int d\vec{r}_3 d\vec{r}_3' \phi_n^*(\vec{r}_3')[1-\eta(\tilde{r}_{13})][1-\eta(\tilde{r}_{23})] \times$$
$$\times W_3(\vec{r}_1' \vec{r}_2' \vec{r}_3' | \vec{r}_1 \vec{r}_2 \vec{r}_3) \phi_n(r_3)[1-\eta(r_{13})][1-\eta(r_{23})]. \tag{9}$$

The function $\eta(r)$ defined as the average over spin and momenta in the Fermi sea of the defect function, where only the most essential partial wave components have been included, i.e., the $^1S_0$ and $^3S_1$ partial waves. The trace is taken with respect to the spin and isospin of the nucleon. In the present work, one may introduce a Skyrme effective interaction density dependent term in addition to the BHF calculation made in Ref. 32 to obtain a correction for the three-body forces.

$$V(\vec{r}_1, \vec{r}_2) = \sum_{i=1}^{4} t_i (1 + x_i P_\sigma) \rho^{\alpha i} \delta(\vec{r}_1 - \vec{r}_2) \tag{10}$$

This is a two-body density dependent potential which is equivalent to three-body interaction. Where $t_i$ and $x_i$ are interaction parameters given in Table I as a result of a least square fit to the exact calculation with three body forces given in Ref. 34, $P_i$ is the spin exchange operator, $\rho$ is the density, $r_1$ and $r_2$ are the position vectors of the particle (1) and particle (2) respectively. $\alpha_i = (1/3; 2/3; 1/2; 1)$. This potential has been used previously by one of the authors Mansour H. [35-39].

## 4. Results and discussion

Now the study of the TBF became necessary in order to modify the failure of two-body forces to achieve the empirical saturation properties of the nuclear matter. Here in this work, we mention only that the calculation of Ref. 32 was performed in the framework a nonrelativistic BHF approach based on the two-body interaction, where the results do not reproduce the empirical point. Then the enhancement came after the addition of TBF's. Therefore, by calculating the correction parameters in the two-body force results in order to achieve the TBF results of Ref. 32 over the whole range of density. These are shown in Table I using a least square fit. The second motivation behind this work is comparing our results for the TBF correction obtained for the symmetric nuclear matter and the pure neutron matter by that found in Ref. 32, using the same BHF approach with the same potentials $AV$18 [6] and the BonnB [11].

In Fig. 1 we show the energy per particle calculated within the scheme BHF vs. the density for the symmetric nuclear matter, (lower curves) for the two-body forces with $AV18$ and BonnB potentials and (upper curves) for the TBF adopted from Ref. 32. We notice that the BHF approach for the two-body forces fail to reproduce the empirical saturation point of the symmetric nuclear matter $(\rho_0 = 0.265\,fm^{-3})$ for the Argonne $V18$ potential and $(\rho_0 = 0.33\,fm^{-3})$ for the BonnB potential. As we can see, the TBF contribution to the EOS of the nuclear matter is repulsive within the BHF framework. The introduction of the TBF shifts and improves the saturation properties of the nuclear matter $(\rho_0 = 0.2\,fm^{-3}, E_0/A = -15.27 MeV)$ for the Argonne $V18$ potential and $(\rho_0 = 0.17\,fm^{-3}, E_0/A = -17 MeV)$ for the BonnB potential) towards to the empirical saturation point. This indeed makes the inclusion of TBF to the EOS calculation very important. Therefore, it was necessary to calculate the correction parameters to be added to the two-body forces. We can see in Fig. 1 when we add the correction part to the two-body forces, then we reproduce the TBF curves in perfect way as shown in the figure.

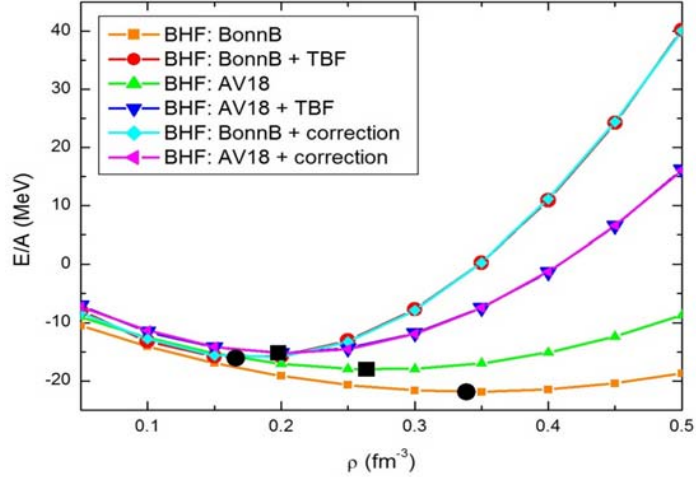

**Fig. 1.** The binding energy per nucleon *E/A* is plotted vs the density for the symmetric nuclear matter, (lower curves) for two-body forces and (upper curves) for the TBF. Using the nonrelativistic BHF calculation with AV18 and BonnB potentials. The last upper two curves show the added corrections to the two-body forces.

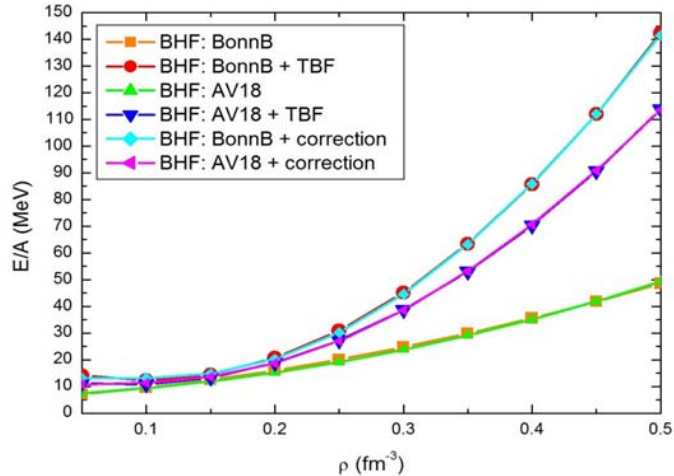

**Fig. 2.** The same as Fig. 1, but for the neuron matter.

The same trend is shown for the pure neutron matter EOS, presented in Fig. 2. It can be concluded that the adopted correction to the two-body forces, both for the symmetric matter and for the pure neutron matter, shows a perfect fitting to the TBF, with different potentials. The study of the symmetry energy ($E_{sym}$) and its dependence of the density is an interesting topic and play an important role in nuclear physics and astrophysics. It is defined from the energy per particle of asymmetric nuclear matter ($\rho_n = \rho_p$ being $\rho_n$ and $\rho_p$ are the neutron and proton densities respectively) as

$$E_{sym}(n) = \frac{1}{2} \left. \frac{\partial^2 \frac{E}{A}}{\partial \beta^2} \right|_{\beta=0}, \quad (11)$$

where $\beta$ is the asymmetry parameter. In this work we adopted the figure of the symmetry energy as a function of density from Ref. 34 trying to apply the correction for the tow-body forces as shown in Fig. 3, the figure shows the symmetry energy in MeV vs. density in $fm^{-3}$ using the BHF for both tow-body forces and the inclusion of TBF illustrating the exact fitting of the correction for both adopted potentials Argone $V$18 and BoonB. At $\rho_0$ the symmetry energy is $32 MeV$.

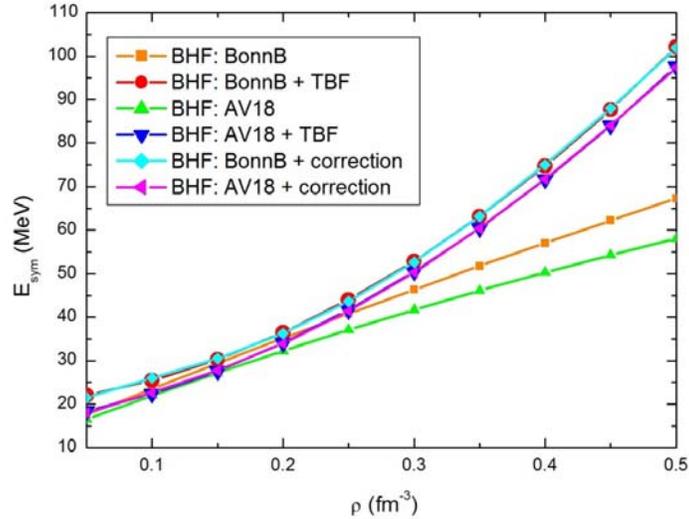

**Fig. 3.** The symmetry energy ($E_{sym}$) in MeV for the EOS using the BHF with and without TBF is plotted vs the density in f m$^{-3}$ with AV18 and BonnB potentials. The fitting of the correction for the two-body force is shown.

The pressure for the nuclear matter can be given from the energy per particle of the symmetric matter as

$$P(\rho) = \rho^2 \left. \frac{dE/A}{d\rho} \right|_A \quad (12)$$

We show in Fig. 4 the pressure in ($MeV.fm^{-3}$) as a function of the density in $fm^{-3}$ or the symmetric nuclear matter. In the same way, we used the $AV18$ and BonnB potentials, and TBF. As we can see clearly, the fitting of the corrections for the tow- body forces along the whole range of the density is very well. Since the figure shows that for the symmetric nuclear matter which is considered bound system, the pressure is negative at densities below the saturation density, gradually the pressure start to increase as the increasing of the density up to zero value at the saturation density $(P=0)$.

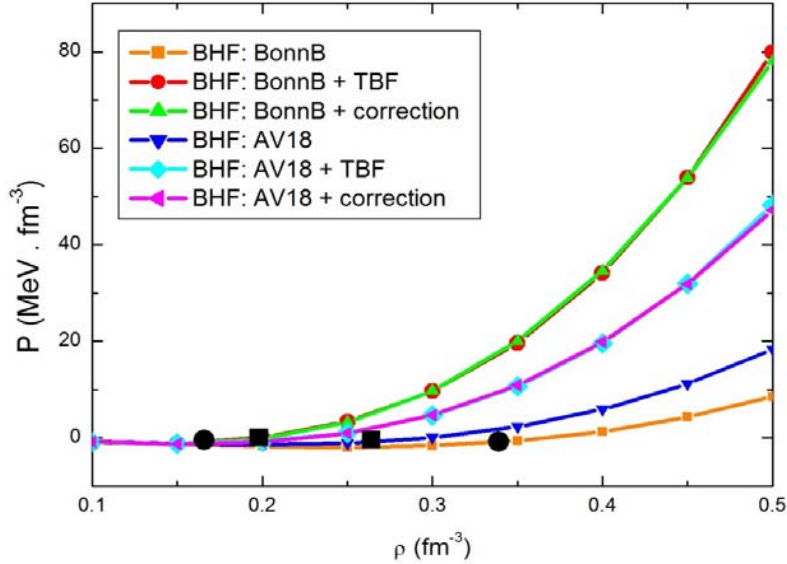

**Fig. 4.** The pressure in (MeV. fm$^{-3}$) for symmetric nuclear matter vs the density in fm$^{-3}$ with AV18 and BonnB potentials.

This explains that the nuclear matter at the saturation density can be in a mechanical equilibrium with external pressure. The nuclear incompressibility $K$ is an important ingredient in the nuclear equation of state. Its value very important into the analysis of astrophysical phenomena such as supernova explosions and heavy ion collisions. It can be used to understand the stiffness of the EOS. Which defined as a slope of the pressure at the saturation density, and it can be calculated from this equation:

$$K = 9\frac{\partial P(\rho)}{\partial \rho}\bigg|_{\rho=\rho_0} = 9\rho^2 \frac{\partial^2 (E/A)(\rho)}{\partial \rho^2}\bigg|_{\rho=\rho_0}. \tag{13}$$

We have calculated the incompressibility at the saturation densities for two different potentials, and it is found 246.97 for $AV18$ and 216.7 for BonnB potentials. Despite of the experimental value of the incompressibility for the symmetric nuclear matter have been determined to be $240 \pm 20 MeV$ [39].

We show in Fig. 5 the incompressibility $K$ as a function of the density using $AV18$ and BonnB potentials with and without the inclusion of the TBF. This figure shows the compatibility between the TBF use and use of the corrections, and reflects the importance of the use of the correction for various densities.

**Table I.** The correction parameters for the two-body forces, using $AV18$ and BonnB potentials.

| Potential | T1 | T2 | T3 | T4 | X1 | X2 | X3 | X4 |
|---|---|---|---|---|---|---|---|---|
| Argonne V18 | 11035 | 31050 | -35125 | -6699 | -12116 | -34450 | 38839 | 7369 |
| BonnB | 18197 | 52141 | -58643 | -11019 | -22080 | -62480 | 70560 | 13500 |

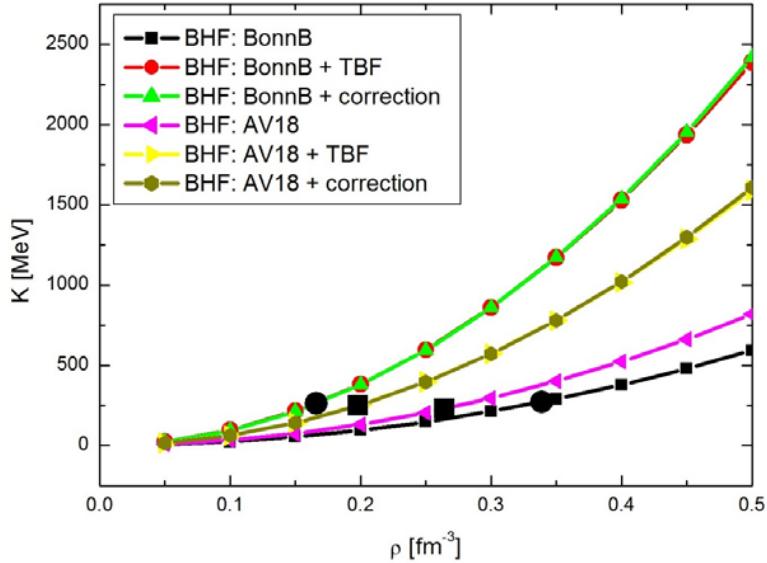

**Fig. 5.** The incompressibility $K$ as a function of the density using AV18 and BonnB potentials with and without TBF.

## 5. Conclusions

Thus, in the present paper we studied the EOS in the framework of the BHF approach following Ref. 32, starting from the symmetric nuclear matter and the pure neutron matter, using the two-body forces and the inclusion of the suggested TBF. We illustrate the correction for the two-body forces, which fit in perfect way the TBF results of Ref. 32. In addition to confirm we have shown figures for the pressure and incompressibility calculated for the same potentials $AV18$ and BonnB used on a wide range of density and we got the expected results with the use of the correction.

**References**

[1] B. D. Day, "Elements of the Brueckner-Goldstone theory of nuclear matter", Rev. Mod. Phys. 39 (1967) 719.
[2] Hessels, Jason; Ransom, Scott M.; Stairs, Ingrid H.; Freire, Paulo C. C.; et al. (2006). "A Radio Pulsar Spinning at 716 Hz". Science, 311 (5769): 1901-1904. arXiv:astro-ph/ 0601337.
[3] Naeye, Robert (2006-01-13). "Spinning Pulsar Smashes Record". Sky Telescope. Retrieved 2008-01-18.
[4] R.B. Wiringa, V. G. J. Stoks, and R. Schiavilla, Phys. Rev. C 51, 38 (1995).
[5] R.B. Wiringa ,R.A. Smith and T.L. Ainsworth, Phys. Rev. C 29 (1984) 1207.
[6] I.E. Lagris and V.R. Pandharipande, Nucl. Phys. A 359 (1981) 349.
[7] R. Machleidt, K. Holinde and Ch. Elster , Phys. Rep. 149 (1987) 1.
[8] V. Stokes and J.J. de Swart , Phys. Rev. C 47 (1993) 761.
[9] I. Bombaci and U. Lombardo, Phys. Rev. C44 (1991) 1892.
[10] I. Bombaci, T.T.S. Kuo and U. Lombardo, Phys. Rep. 242 (1994) 165.
[11] R. Machleidt, Nucl. Phys. 19, 189 (1989).
[12] J.P. Jeukenne, A. Lejeune, C. Mahaux, Phys. Rep. 25, 83 (1976).
[13] F. Coester, S. Cohen, B.D. Day, C.M. Vincent, Phys. Rev. C 1 (1970) 769.
[14] M. Baldo (Ed.), Nuclear, Methods and the Nuclear Equation of State, International Review of Nuclear Physics, vol. 8, World Scientific, 1999 (Chapter 1).
[15] H.Q. Song, M. Baldo, G. Giansiracusa, U. Lombardo, Phys. Rev. Lett. 81 (1998) 1584.
[16] M. Baldo, G. Giansiracusa, U. Lombardo, H.Q. Song, Phys. Lett. B 473 (2000).
[17] A. Akmal, V.R. Pandharipande, D.G. Ravenhall, Phys. Rev. C 58 (1998) 1804.
[18] H.R. Moshfegh and M. Modarres, Nucl. Phys. A792, 201 (2007).
[19] M. Baldo and H.R. Moshfegh, Phys. Rev. C86 024306 (2012).
[20] Wiringa R.B., Fiks V. and Fabrocini A., 1988, Phys. Rev. C38, 1010.
[21] B. S. Pudliner, V. R. Pandharipande, J. Carlson et. al. Phys. Rev. Lett. 74, 4396 (1995).
[22] M. Baldo and L. S. Ferreira, Phys. Rev. C 59, 682 (1999).
[23] V. Soma, P. Bozek, Phys. Rev. C 78, 054003 (2008).
[24] V. Soma, P. Bozek, Phys. Rev. C 80, 025803 (2009).
[25] J. Carlson, R. Schiavilla, Rev. Mod. Phys. 70, 743 (1998).
[26] S.C. Pieper, R.B. Wiringa, J. Carlson, Phys. Rev. C 70, 054325 (2004).
[27] P. Grange, A. Lejeune, M. Martzolff, J.F. Mathiot, Phys. Rev. C 40, 1040 (1989).
[28] W. Zuo, A. Lejeune, U. Lombardo, J.F. Mathiot, Nucl. Phys. A 706, 418 (2002).
[29] Schiavilla R., V.R. Pandharipande and Wiringa R.B., 1986, Nucl. Phys. A449, 219.
[30] M. Baldo, Alaa Eldeen Shaban, Phys. Lett. B 661 (2008) 373377.
[31] X.R. Zhou, G.F. Burgio, U. Lombardo, H.-J. Schulze, W. Zuo, Phys. Rev. C 69, 018801 (2004).
[32] W. Zuo, I. Bombaci, and U. Lombardo, Eur. Phys. J. A (2014) 50: 12.
[33] H. M. M. Mansour and A. Gamoudi, Physics of atomic nuclei 75, 430(2012).
[34] Hesham M. M. Mansour, Khaled S.A.Hassaneen, Journal of Nuclear and Particle Physics 2(2)(2012) 14.
[35] Khaled Hassaneen and Hesham Mansour, Journal of modern physics 4(5B), 37(2013).
[36] Khaled Hassaneen and Hesham Mansour, Journal of Nuclear and Particle Physics, 2013, 3(4):77-66.
[37] H.M.M. Mansour and Kh.S.A. Hassaneen, Physics of atomic nuclei, 2014, vol. 77, No. 3, pp. 290-298.
[38] P. E. Haustein, Atomic Data and Nuclear Data Tables 39, 185 (1988).
[39] J.R. Stone, N.J. Stone and Moszkowski S.A., Phys. Rev. C 89 (2014) 044316.